# Novel Nanoscroll Structures from Carbon Nitride Layers

Eric Perim and Douglas S. Galvao
Instituto de Física 'Gleb Wataghin', Universidade Estadual de Campinas, 13083-970, Campinas, São Paulo, Brazil.

## ABSTRACT

Nanoscrolls consist of sheets rolled up into a papyrus-like form. Their open ends produce great radial flexibility, which can be exploited for a large variety of applications, from actuators to hydrogen storage. They have been successfully synthesized from different materials, including carbon and boron nitride. In this work we have investigated, through fully atomistic molecular dynamics simulations, the dynamics of scroll formation for a series of graphene-like carbon nitride (CN) two-dimensional systems: g-CN, triazine-based (g-$C_3N_4$), and heptazine-based (g-$C_3N_4$). Carbon nitride (CN) structures have been attracting great attention since their prediction as super hard materials. Recently, graphene-like carbon nitride (g-CN) structures have been synthesized with distinct stoichiometry and morphologies. By combining these unique CN characteristics with the structural properties inherent to nanoscrolls new nanostructures with very attractive mechanical and electronic properties could be formed. Our results show that stable nanoscrolls can be formed for all of CN structures we have investigated here. As the CN sheets have been already synthesized, these new scrolled structures are perfectly feasible and within our present-day technology.

## INTRODUCTION

Nanoscrolls are nanostructures that consist of layered structures rolled up into papyrus-like form [1]. Graphene, carbon nanotubes (CNTs), and carbon nanoscrolls (CNSs) (Figure 1) can be considered, from a topological point of view, as closely related structures, as tubes and scrolls can be formed from rolling up graphene sheets. In fact, scroll can be considered as sheets rolled up into Archimedean spirals [1]. CNTs and CNSs differ only by the fact the CNSs present both ends open, since the edges of the scrolled membranes are not fused.

CNSs are remarkable structures sharing some of the graphene and CNTs properties, and also exhibiting unique ones [2,3]. Due to their open topology their diameter can be easily tuned, which can be the basis for a large variety of applications, from hydrogen storage [4] to electromechanical nanodevices [5], among others [1].

CNSs have a long history, with their first observation dating back to 1960s [6]. Difficulties in synthesis and characterization [7] have precluded their wide investigations. Recent advances in synthesis [8] have renewed the interest in these nanostructures.

Similarly to CNTs, nanoscrolls of different materials are possible. In fact hexagonal boron nitride nanoscrolls were theoretically predicted [9], and recently experimentally realized [10]. In principle, any layered structure is a good candidate for scrolls, as the mechanisms for the scroll formation and stability is an interplay between van der Waal interactions (energetic gain associated with the overlap regions) and mechanical deformations (energetic cost associated with bending the layers). In order to overcome the barrier to scroll formation, some energy should be provided to the system [2,3].

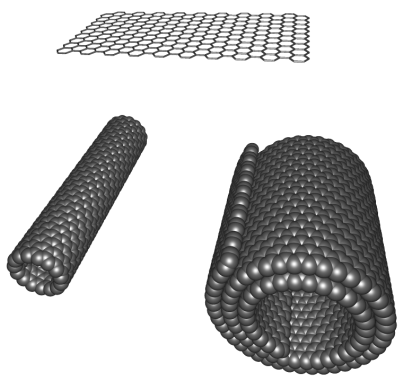

**Figure 1.** Graphene (top), carbon nanotubes (left), and carbon nanoscrolls (right). See text for discussions.

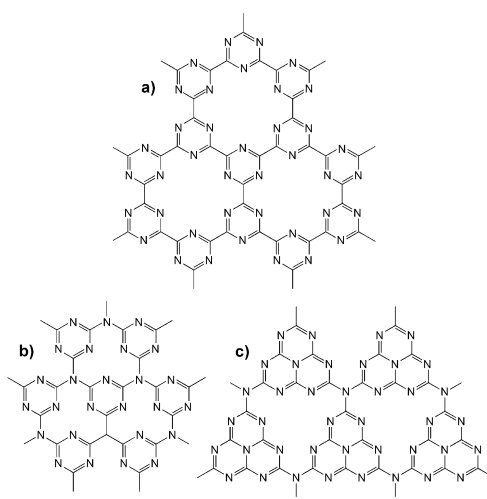

**Figure 2.** Graphene-like CN layer structures: a) g-CN; b) triazine-based g-$C_3N_4$, and; c) heptazine-based g-$C_3N_4$.

In this work we have investigated, through fully atomistic molecular dynamics simulations, the dynamics of scroll formation for a series of graphene-like carbon nitride (CN) two-dimensional systems: g-CN, triazine-based (g-$C_3N_4$), and heptazine-based (g-$C_3N_4$) (Figure 2). Carbon nitride (CN) structures have been attracting great attention since their prediction as super hard materials. Recently, several graphene-like carbon nitride (g-CN) structures have been synthesized with distinct stoichiometry and morphologies [11].

**METODOLOGY**

We carried out fully atomistic molecular dynamics (MD) simulations to investigate the structural and dynamical aspects of scroll formation for the structures presented in Figure 2. The simulations were carried out using the universal force field (UFF)[12], as implemented in the Materials Studio suite [13].

The MD simulations were carried out using a NVT ensemble, i. e., constant volume, temperature, and number of atoms. The used temperature of 50 K was controlled using a Nose-Hoover thermostat [12,13], with time steps of 1 fs. All atoms were assumed to be in neutral state and with partial double bonds. No charge optimization processes were used.

The structural parameters for the rolling up scroll analyses are indicated in Figure 3. These parameters are defined by their parent (unscrolled planar configuration) sheet dimensions (L and W), and by the scrolling angle θ.

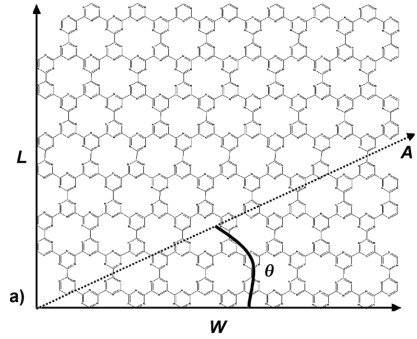

**Figure 3.** Scheme and the structural scroll parameters.

**DISCUSSION**

As mentioned above, the balance between van der Waals and elastic forces defines the formation and scroll stability. Starting from a planar configuration, bending the structure increases the energy system, which requires an energy assisted process or the system will return to its planar configuration. However, if the bending process goes beyond a critical point where there is an overlap between layers, the van der Waals forces (in opposition to the elastic ones) increases the stability of the system and depending on the overlapping area value, this could lead to a spontaneous and self-sustained scroll formation.

This energy balance can be better analyzed defining the quantity ΔE, that is the total energy value minus the energy of the planar configuration (reference energy), divided by the number of atoms in the structure. In Figure 4 we present these results for g-CN nanoscrolls, for the case of W=L=100 Å, and θ=0°. For comparison purposes the data for equivalent carbon nanoscrolls are also displayed.

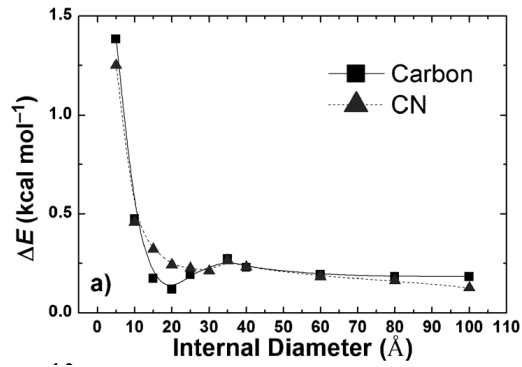

**Figure 4.** ΔE as a function of the internal diameter values for different carbon and g-CN scrolls.

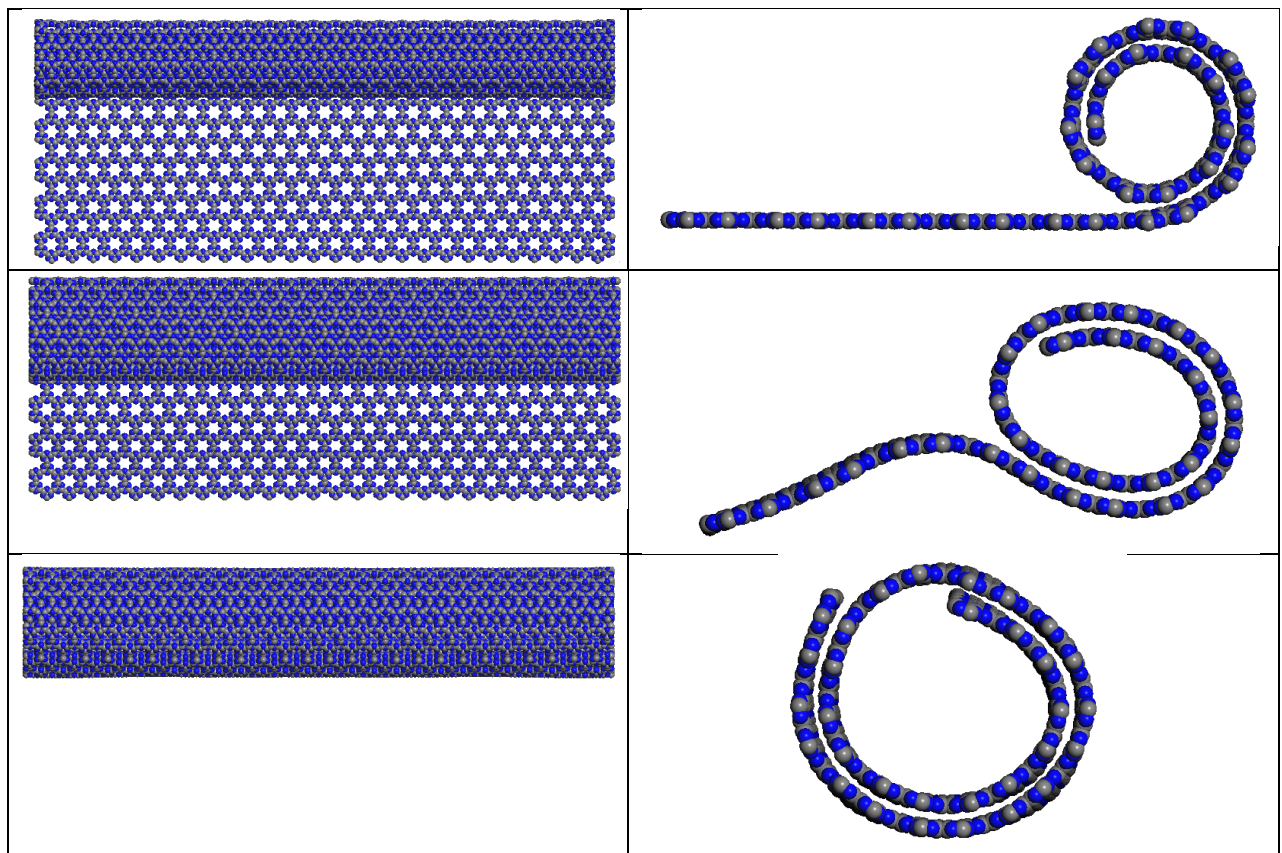

**Figure 5.** Snapshots from MD simulations (frontal and cross-section views) of initial, intermediate and final stages of the g-CN scroll formation.

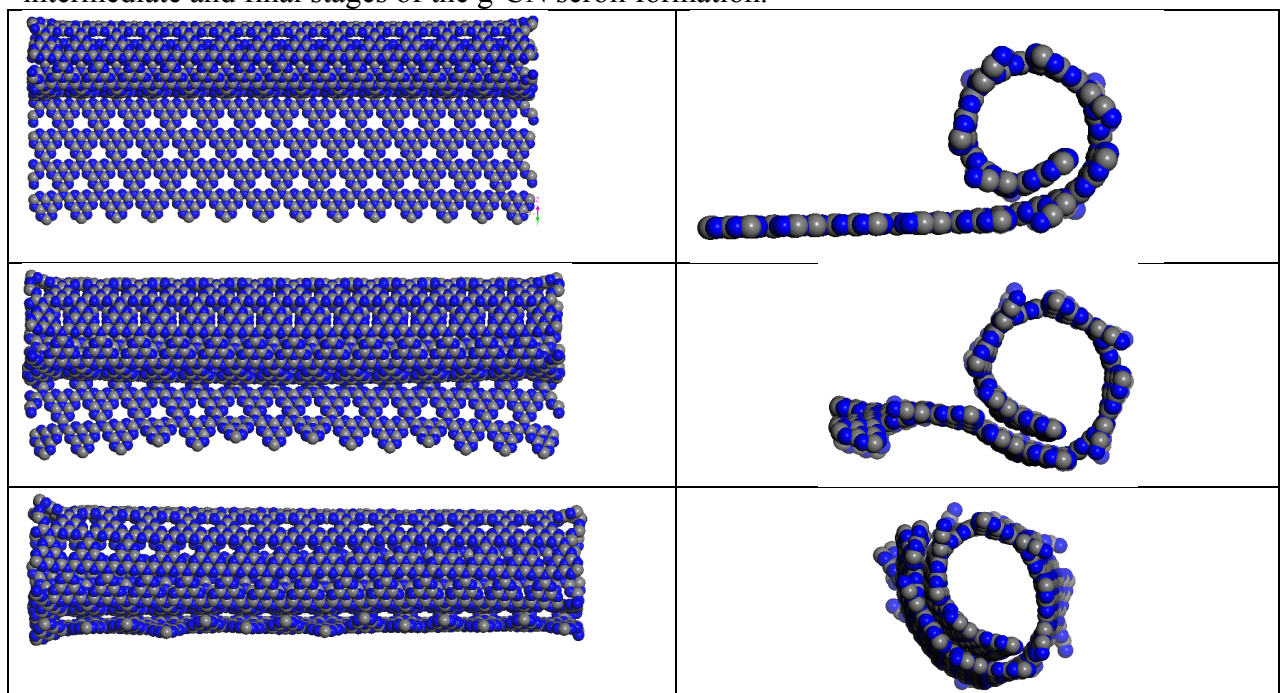

**Figure 6.** Snapshots from MD simulations (frontal and cross-section views) of initial, intermediate and final stages of the heptazine-CN scroll formation.

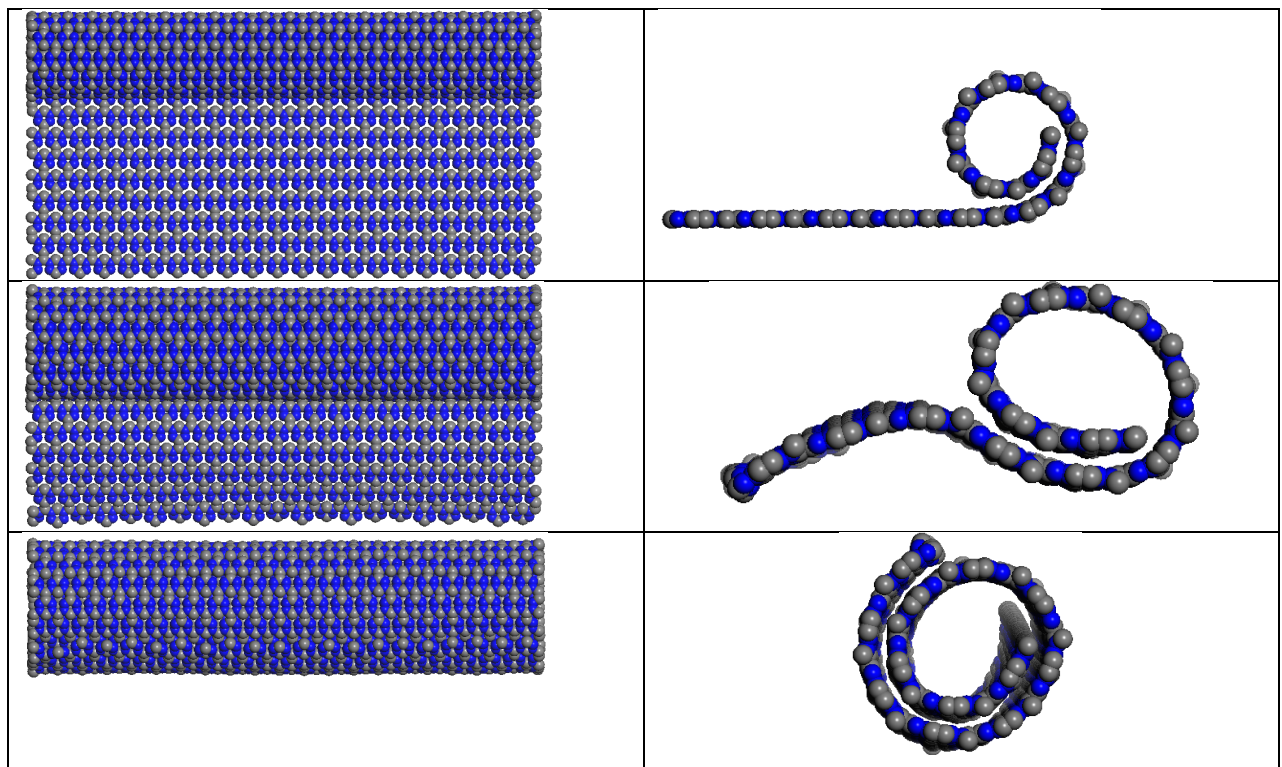

**Figure 7.** Snapshots from MD simulations (frontal and cross-section views) of initial, intermediate and final stages of the triazine-CN scroll formation.

As can see from this Figure there is a well-defined minimum around 20 Å, which indicate the region where the scroll can be formed and be stable. This minimum is even deeper than for the case of the carbon nanoscrolls. This can be explained by the fact that g-CN sheets are largely porous (not the case of graphene), which results in a smaller contact area and decreased interlayer interactions [14]. Similar results were obtained for triazine and heptazine CN-scrolls.

We have also investigated how the scroll stability depends on L and W values. Our results showed that as W increases, also increases the scroll stability. This is a direct consequence that the number of overlap layers is proportional to W, resulting in a stronger interlayer interactions. However, the L values have almost no effect on the scroll stability, except for extremely small L values. This implies that the scroll length does not significantly affect its stability.

Our results also showed that stable nanoscrolls could be formed for all of CN structures we have investigated here. In Figures 6-8 we present snapshots from MD simulations showing different stages (initial, intermediate and final) of the scroll formation for g-CN, heptazine-CN, and triazine-CN structures, respectively.

**CONCLUSIONS**

In summary, we have investigated using fully atomistic molecular dynamics simulations the structural and dynamical aspects of scroll formation for a series of graphene-like carbon nitride (CN) two-dimensional systems: g-CN, triazine-based (g-$C_3N_4$), and heptazine-based (g-$C_3N_4$).

Our results show that stable nanoscrolls could be formed for all of CN structures we have investigated here. The diameter value stability is around 20 Å. Our results showed that as W (see

Figure 3) increases, also increases the scroll stability, while the L values have almost no effect on the scroll stability, except for extremely small L values. As the CN sheets have been already synthesized, these new scrolled structures are perfectly feasible and within our present-day technology.

## ACKNOWLEDGEMENTS

This work was supported in part by the Brazilian Agencies CAPES, CNPq and FAPESP. The authors thank the Center for Computational Engineering and Sciences at Unicamp for financial support through the FAPESP/CEPID Grant # 2013/08293-7.

## REFERENCES

1. E. Perim, L. D. Machado, and D. S. Galvao, *Frontiers in Materials* **1**, 31 (2014).
2. D. Tomanek, Physics B **323**, 86 (2002).
3. S. F. Braga, V. R. Coluci, S. B. Legoas, R. Giro, D. S. Galvao, and R. H. Baughman, *Nano Lett.* **4**, 881 (2004).
4. V. R. Coluci, S. F. Braga, R. H. Baughman, and D. S. Galvao, *Phys. Rev. B* **75**, 125404 (2007).
5. R. Rurali, V. R. Coluci, and D. S. Galvao, *Phys. Rev. B* **74**, 085414 (2007).
6. R. Bacon, *J. Appl. Phys.* **31**, 283 (1960).
7. L. M. Viculis, J. J. Mack, and R. B. Kaner, *Science* **299**, 1361 (2003).
8. J. Zheng *et al.*, *Adv. Mater.* **23**, 2460 (2011).
9. E. Perim and D. S. Galvao, *Nanotechnology* **20**, 335702 (2009).
10. X. Chen, R. A. Boulos, J. F. Dobson, and C. L. Raston, *Nanoscale* **5**, 498 (2013).
11. J. Li, C. Cao, and H. Zhu, *Nanotechnology* **18**, 115605 (2007).
12. A. K. Rappe, C. J. Casewit, K. S. Colwell, W. A. Goddard III, and W. M. Skiff, *J. Amer. Chem. Soc.* **114**, 10024 (1992).
13. Materials Studio suite is available from Accelrys. http://www.accelrys.com.
14. E. Perim and D. S. Galvao, *ChemPhysChem* **15**, 2367 (2014).